\documentclass[english,aps,prstper,reprint,showpacs,titlepage,longbibliography]{revtex4-2}   

\usepackage[T1]{fontenc}	
\usepackage[latin9]{inputenc}	
\usepackage{geometry}		
\usepackage{graphicx}
\usepackage[above,below]{placeins}	
\usepackage{times}

\usepackage{hyperref}  
\hypersetup{colorlinks=true,urlcolor=blue,citecolor=blue,linkcolor=blue}   
\usepackage{enumitem}          
\setlist{nosep}                 

\usepackage{graphicx}
\usepackage{subcaption}
\usepackage{dcolumn}
\usepackage{tabularx}
\usepackage{multirow}
\usepackage{bm}
\usepackage{amsfonts,color} 
\usepackage[table,xcdraw]{xcolor}
\usepackage[english]{babel}
\usepackage{lipsum} 
\usepackage{array}

\definecolor{Gray}{gray}{0.9}

\newcolumntype{Y}{>{\centering\arraybackslash}X}
\newcolumntype{Z}{>{\raggedright\arraybackslash}X}

\newcolumntype{P}[1]{>{\raggedright\arraybackslash}p{#1}}

\begin{document}

\begin{titlepage}

  \title{Inequities and misaligned expectations in PhD students' search for a research group}
 
\author{Mike Verostek (he/him)}
\affiliation{
 Department of Physics and Astronomy, University of Rochester, Rochester, New York 14627 
}
 \affiliation{School of Physics and Astronomy, Rochester Institute of Technology, Rochester, New York 14623}

\author{Casey W. Miller (he/him)}
\affiliation{
 School of Physics and Astronomy, Rochester Institute of Technology, Rochester, New York 14623
}

\author{Benjamin M. Zwickl (he/him)}
\affiliation{
 School of Physics and Astronomy, Rochester Institute of Technology, Rochester, New York 14623
}


  \begin{abstract}
    Joining a research group is one of the most important events on a graduate student's path to earning a PhD, but the ways students go about searching for a group remain largely unstudied. It is therefore crucial to investigate whether departments are equitably supporting students as they look for an advisor, especially as students today enter graduate school with more diverse backgrounds than ever before.  To better understand the phenomenon of finding a research group, we use a comparative case study approach to contrast important aspects of two physics PhD students' experiences.  Semi-structured interviews with the students chronicled their interactions with departments, faculty, and the graduate student community, and described the resources they found most and least helpful.  Our results reveal significant disparities in students' perceptions of how to find an advisor, as well as inequities in resources that negatively influenced one student's search.  We also uncover substantial variation regarding when in their academic careers the students began searching for a graduate advisor, indicating the importance of providing students with consistent advising throughout their undergraduate and graduate experiences.

    \clearpage
  \end{abstract}
  

  \maketitle
\end{titlepage}

\section{\label{sec:Introduction}Introduction and Background}
\vspace{-2mm}

Current data indicates that the retention rate of physics PhD students is approximately 60\%, with attrition from PhD programs disproportionately affecting traditionally underrepresented students \cite{cgs2008phd, miller2019typical, lott2009doctoral}.  Leaving a PhD program can adversely affect students' mental health and financial well-being, and early departures impact graduate programs that must invest time and resources into recruiting and supporting new students \cite{lovitts2002leaving, golde2005role}.  As the number of physics PhDs granted across the US grows and the population of physics graduate students becomes more diverse than ever before \cite{aip2021trends}, investigating the underlying factors behind high attrition is imperative. 

Previous studies on graduate attrition across STEM and non-STEM disciplines indicate that a negative advising relationship is a key factor that motivates students to leave \cite{devos2017doctoral, lovitts2002leaving, rigler2017agency, golde1998beginning, jacks1983abcs}.  At present, much research on graduate advising relationships has sought to identify which qualities are indicative of productive mentorships \cite{schlosser2003qualitative, schlosser2011multiculturally, barnes2010characteristics, bargar1983advisor}. Despite the fact that finding an advisor is often cited as a crucial decision for PhD students to make \cite{bloom1999ph, barres2013pick}, existing studies generally do not examine the process by which these relationships form.  PhD students often find navigating their first year of study to be difficult \cite{gardner2007heard}, and if the process of finding a group is difficult for students to navigate, they may be less likely to find a group that provides them with a fulfilling research experience.  

In one survey of approximately 4,000 STEM and non-STEM PhD students, students who reported being satisfied with their advisor tended to take more factors into account when looking for a research group than students who reported advisor dissatisfaction \cite{golde2001cross}.  This suggests students' access to information while searching for research groups is important in their eventual satisfaction in graduate school.  However, it remains unclear how students go about gathering information about groups, what they prioritize, and which resources they find most helpful in that process.  It is also unclear if there are differences across student demographics.

We investigate the phenomenon of finding a research group in the specific context of physics PhD programs using a comparative case study approach.  By contrasting two cases of physics PhD students searching for a graduate research group, we sought to answer the overarching research question of, ``How do the experience and process of seeking and finding a research group vary among physics PhD students?''  Within this overarching question, we also asked the following subquestions: ``What are some of the major questions and concerns that students have while looking for a research group?'' and ``What helps students answer these questions, and what gets in their way?''



\vspace{-2mm}
\section{\label{sec:Method}Method}
\vspace{-2mm}

Case studies leverage specific cases to generate in-depth illustrations of an issue or problem \cite{creswell2016qualitative}.  Here, the cases for comparison are the narratives of two physics PhD students who recently completed their search for a graduate research advisor.  Although case studies typically involve collection of data from multiple sources (e.g., artifacts, direct observation), our analysis is solely based on interviews with the two students.  These were conducted as part of a larger study that includes over 40 interviews and is intended to characterize how students experience the process of finding a research group.  However, due to the limited space available we opted to select two particular students with divergent perspectives on searching for an advisor and analyze them separately from the others. These two cases vividly demonstrate the impact that differing levels of access to resources can have on students' graduate experience, thereby highlighting several important ways departments must better support students during their search for a research group.

Our semi-structured interview protocol was inspired by cognitive task analysis (CTA) methods \cite{crandall2006working, clark2008early} and Dervin's sense-making method \cite{dervin1986neutral, dervin2003sensemaking}.  These methodologies are designed to elicit detailed descriptions of interviewees' thoughts and actions as they recount how they progressed toward a goal, in this case joining a research group.  Moreover, CTA and sense-making focus on specific life experiences that are bounded in time.  These features make our data amenable to a case study analysis approach, since cases must have clear boundaries and great depth.

The protocol is broadly broken into three stages.  First, we gathered a timeline of \textit{Steps} that students took in their search for a research group.  
Students defined the start and end points of their stories, so these boundaries varied based on each student's individual experience.  Examples of common steps included ``Applying to graduate school'' or ``Attending visiting weekend.''  We then asked students to go through each step and identify their major \textit{Questions} regarding their search for a research group at that point in time.  This allowed us to gain insight into which aspects of the process drove confusion and uncertainty.  
Lastly, we asked students to identify any sources of \textit{Help} that allowed them to resolve their question or concern, as well as any obstacles that \textit{Hurt} their ability to move forward.  This stage allowed us to understand the thoughts, actions, and events that helped or hurt students' ability to navigate their search for a group.  After interviews were completed, they were transcribed and edited for grammar and clarity.

Next, focusing on only the two cases for this analysis, we coded the two students' transcripts using a priori codes that aligned with the stages of the protocol: Steps, Questions, Helps, and Hurts.  By associating Questions with particular Steps and Helps/Hurts with particular Questions, we generated timelines of each student's individualized experience.  From these timelines as well as repeated reading of the original transcripts, we wrote narrative summaries of the two students' stories.  These narratives helped us to begin identifying themes within each case, and eventually formed the basis for the case descriptions reported in the results.  
Lastly, we contrasted the Questions, Helps, and Hurts between the two students.  This allowed us to make specific comparisons between each student's sources of confusion, as well as the resources they used to navigate their advisor search.  These cross-case analyses allowed us to generate several general observations and takeaways, which we discuss in Section \ref{sec:Discussion}.   


\section{\label{sec:Results}Results}
\vspace{-2mm}


This section describes the cases of PhD students Alex and Brianna (both pseudonyms).  Alex is a first year physics PhD student at a public doctoral university with Very High research activity.  He self-identifies as Male and Hispanic.  Alex attended an elite private undergraduate institution, and one of his parents holds a PhD in a STEM field.  Brianna is a second year physics PhD student at a public doctoral university with Very High research activity.  She self-identifies as Female and Black, and noted that she was the first in her family to attend graduate school.  Her undergraduate degree was from a large public institution with Very High research activity.  These two PhD students had drastically different experiences joining a research group in graduate school.  For Alex, ``things worked fantastically well,'' as he matched with his advisor and began research the summer before graduate school started.  On the other hand, Brianna indicated that ``the process for me was pretty rough,'' and her search for an advisor lingered into her second year of graduate school.  Alex and Brianna's individual narratives will provide a basis for identifying factors that led to their divergent experiences. 

It is notable that COVID restrictions undoubtedly had some impact on both students' their experiences.  The strongest restrictions were in effect during Alex's junior and senior undergraduate years, and Brianna's senior undergraduate year and first year of graduate school.  However, neither student cited COVID as a major source of adversity, only noting that some restrictions exacerbated existing issues.


\textbf{Case 1 (Alex):} For Alex, the search for a graduate research group began the summer following his undergraduate junior year when he participated in an REU at his top-choice graduate school.  Being in this position gave him access to graduate students and prospective PIs with whom he could conduct lab tours. ``I basically just asked the grad student I was working with on the REU `Hey, do you know anyone in these two groups?' And I looked them up and emailed them and said, 'Hey, I'm a student, I will be applying to grad school next year, I like your labs. Do you have time to show me around?'''  Following lab tours with graduate students, he recalled following up via email with the lab PIs, telling them ``I am a student that will be applying that's interested in your group. I did a tour with ``name,'' and I really liked ``project.'' Can we meet to talk further about whether or not you'll be hiring students? My CV is attached.''  Alex reported that during the summer, ``I also met with the chair of the physics department and said `I really want to come here... Advice on applying effectively?' And you know, once you get your foot in the door with the whole, like, I'm [a] physics REU student, the doors kind of open to you.'' Alex said that his REU allowed him to ``tailor'' his application to his top school.

Alex said that he ``screened" potential graduate schools by research interests, looking to align his interest in experimental optics with groups at each school.  
He noted that ``I got the advice about applying to places from my advisor saying, be persistent, make connections, network, do everything you can before the application goes in to make sure they know your name.''  Alex was admitted to his top choice school, and expressed that at that time he felt ``relaxed" because ``I have a place to go. I like the people there.  They want me to work for them."
During visiting weekend, Alex said that while most students wanted to ``decide if they wanted to come to [the school]," he was spending his time narrowing down his advisor search.  He again received advice from his undergraduate faculty advisor, this time regarding what to look for in prospective research groups: ``I asked them, `Hey, what should I keep an eye out on for red flags?''  Moreover, the access he received as an REU student allowed him to more easily to reach out to professors at the school a second time; in some cases, professors reached out to him.  
``That's where I got to ask about work life balance, attrition rate, flexibility with choosing mentors. Are there rotations or are there not rotations? What does funding look like? How big are my classes? Can I take classes outside of physics? What do students do?''  He noted that ``It helped that I'm not the first person in my family to go to graduate school'' when he was coming up with questions to ask PIs.  He also asked again whether each PI was planning to hire students in the upcoming year, although Alex's funding concerns were eliminated when was awarded a graduate fellowship.  

Having the advantage of being able to meeting professors during his senior year meant that Alex felt he had ``so much time" to schedule meetings with graduate students, who he said he considered a vitally important part of his search process.  In meetings with graduate students, he asked questions about lab culture, advisor support, work life balance, friends, classes, and general quality of life. ``All of those questions, the culture and the like, were answered when I literally met with the grad students. I spent two and a half, three hours going through... I mean that's why visiting is so important. But that's also why having months instead of just a couple of days, where I could plan with so much time, if I have more questions I can come back in a few days.''  

Alex was particularly interested in two research groups at his chosen school, and he described how ``universally positive" feedback from both graduate students and the PIs made him ``comfortable" with wherever he chose to go.  
Ultimately, he was ``sold" on one of the projects when a PI told him, ``If you're also excited about doing this, which it sounds like you are, then let's do it together... Like he basically said, `You are my first choice to work on this project.'''  That advisor also went to Alex's undergraduate institution for his postdoc, which contributed to their positive connection.


Reflecting on his experience seeking out a research group, Alex suggested that his experience might be ``atypical'' and ``somewhat idealistic.'' 
Overall, he said that he often felt as though departments ``leave the student to fend for themselves.  You figure it out.  We'll help you if you ask, but we're not really going to set up structures that promote your ability to efficiently meet with laboratories and students.''  Despite this, Alex attributed his perceived success in the process to ``find[ing] someone in a key position'' to help ``move the ball forward.''
  

\textbf{Case 2 (Brianna):} In selecting where to attend graduate school, Brianna highly valued her institution's proximity to family and its superior financial package. She recalled, ``I originally came to grad school with the idea that I'd be doing research in astrophysics. And I chose [this school] not because of the research, but because of the proximity... I saw that they did have a good like three or four astrophysicists here, so I thought I'd be able to do research in astrophysics.''  However, upon arriving in her first year, Brianna reported  difficulty navigating through graduate school: ``Coming into grad school, I didn't know anything about grad school. I'm the first in my family to go to grad school. So I had no idea what I was doing.''    

In particular, Brianna felt that her physics department ``didn't really place an emphasis on finding a research advisor.''  
She remembered feeling unclear as to ``whether or not I had to find [an advisor] right away, as opposed to like, waiting out and focusing on trying to pass the courses first. Because I feel like doing all of that was a little bit stressful.''  
Brianna reported feeling concerned that she might be falling behind, but also felt ``too scared" to ask for guidance from faculty who ``hold the fate of your career in their hands.''  She remembered thinking that she ``just didn't know what was appropriate to ask and what not to ask. Like, I don't know what I should have known coming in.''  Meanwhile, Brianna said the pandemic had exacerbated her feelings of isolation from other graduate students, who she felt would have helped her ``put things in perspective."   This prompted her to look online for advice, recalling how, ``Honestly, I would just go onto grad students subreddits and be like, okay, is this normal? Because it was during the pandemic, so I couldn't really talk to anybody.''  Although some online resources were helpful, she said they more frequently made her ``[feel] bad'' because students online ``have everything set up and they've only been in grad school a month... like, I don't know why I'm not as far along as everybody else.''

Shortly before the end of Brianna's first year the department-level faculty advisor informed her that she needed to find a research group by the start of the summer, which was an unexpected development.  ``It was kind of blindsiding me because the general student advisor had initially said don't worry about it, you have plenty of time to find a research advisor... I was under the impression that could wait until my second year.'' 
At that moment, Brianna said she felt ``scared'' and ``taken aback.''  She then set up a meeting with one of the astrophysics professors in the department, but he told her that he had taken on a new student just the day before, and was unable to take another.  

In June of the first year summer she reached out to another astrophysics professor who informed Brianna that he might have funding to take on another student, but was not certain.  Although this faculty member had neither accepted her nor supported her financially yet, she described feeling ``bound'' to his group and said did know if she could talk to other potential advisors at the same time.  She explained,  ``I didn't know if I was allowed to reach out to other people during the time waiting for him to come back to me.''  In September, the professor told her that he did not have funding. ``In the most plain terms he just kind of wasted my time for like, three or four months, when I could've used that time to talk to other professors. So I ended up talking to two more astrophysics professors asking if I could be in their group. And both of them said no, because they didn't have funding for another grad student. So that's all four in the department.''  Having exhausted all of her options in the school's astrophysics community, Brianna was uncertain about whether she should stay in the department.

At the time, Brianna felt her difficulty finding an advisor was due to a fault of her own, saying ``Personally, I thought I was cursed. Like I slighted the department in some way, thinking that I had offended someone and I was being blacklisted from the research advisors. It didn't make any sense. And I was just really having a hard time for a minute because I was like, I don't know what I'm going to do now, if I can't stay in the program.''  By then, she was aware that many of her peers had found advisors, which added to her feelings of isolation, saying ``I'm suffering alone. And I'm too embarrassed to admit it.''  Brianna reflected that her journey in graduate school had been ``just really depressing at points...
It was like, I guess I'm not supposed to be here. Because everything is making it seem like it's not working out.''  

In what Brianna described as one last ``desperate plea,'' she approached her former E\&M professor to ask if he would be able to advise her.  She figured that she could ``learn to like'' some part of his research, but maintained worry that would not be the case.  Still, she joined his group, motivated to persevere in her PhD.  ``He was like, `Yeah, we'll be able to find something for you. You might be a TA a little bit longer, but we'll figure something out for you.' And he did his best to give me a project that was still kind of astrophysics related.''  Brianna is currently still in his group and believes his willingness to adjust his own research agenda to better fit her interests has allowed them to work well together.

Brianna summed up her experience by saying ``I was under the impression that I'd get a little bit more advising help, like how to navigate through grad school. And like I said, I really just had to figure it out on my own.''  She elaborated, ``I think I made the mistake of trying to place my trust into the professors, when I probably should have been trying to make connections with other grad students. Because they know what's going on. And if you're able to successfully do that, I feel like then they can actually tell you the truth.''

\vspace{-2mm}
\section{\label{sec:Discussion}Discussion}
\vspace{-2mm}

Despite being largely absent from prior research, these cases demonstrate the significant ramifications the group search process can have for new graduate students and indicate that the search process itself may be an important source of graduate attrition.  While Alex described feeling like advisors ``want[ed] me to work for them,'' being rejected by several groups made Brianna feel ``cursed,'' ``alone,'' and that ``I'm not supposed to be here.''  Rejection from potential research groups, perceived by Brianna as a personal failing, clearly influenced Brianna's sense of belonging and drove her to consider leaving the program.  Not only that, but the numerous advantages Alex enjoyed while navigating the search process allowed him to begin conducting research a full year before Brianna, potentially shortening his overall PhD timeline, and resulted in his ability to choose the exact research lab he wanted to join.  

Due to the highly impactful nature of the group search process, it is critical for students of all backgrounds to be supported in their group search, and it is incumbent upon departments to provide that support.  Indeed, we wish to emphasize that the cases presented here are not meant to establish guidelines for what \textit{students} should do while searching for a group, but rather to highlight that \textit{departments} must better support students during their search for a research group.

\textbf{Departments must create new structures to clarify expectations and guide students in their group search:}

Both Alex and Brianna reported that they did not feel their departments provided guidance and support when it came to finding research groups.  However, Alex's background as an REU participant, student at an elite undergraduate institution, and son of an advanced degree holder allowed him to navigate the lack of structure with his incoming knowledge of the system.  Meanwhile, Brianna's background left her feeling like she ``didn't know anything about grad school" and constantly wishing for formal guidance about her timeline, where to seek help, and norms of communication.  Departments that lack structures to accommodate students whose backgrounds resemble Brianna's therefore fundamentally favor the success of students more similar to Alex.

Recognizing that students enter graduate school with a diverse array of prior experiences, departments must implement programs to give all students the best chance of entering into a fulfilling advising relationship.  Although detailed recommendations are beyond the scope of this paper, the inequities revealed between Alex and Brianna's cases begin to provide insight into the ways departments must improve to make this happen.  For example, departments must strive to provide clearly accessible sources of information that elucidate expectations for students.  This might include information such as when in the PhD students are expected to join a group and where funding is expected to come from, both aspects of the process that Alex was able to anticipate but Brianna expressed difficulty figuring out.




\textbf{Graduate coursework should be concurrently reformed to more strongly emphasize finding a group:}

Research comprises the majority of the graduate school experience, which would seemingly make the process of joining a group a top priority for first year students.  Yet Brianna felt conflicted as to whether to focus on doing coursework or finding an advisor, and cited coursework as actively hindering her ability to seek out a research group because of the amount of time she dedicated to classes.  Although the purpose of graduate coursework is ostensibly to prepare students for research, Brianna's case indicates that the present emphasis on coursework in the first year can actually hurt students' ability to figure out what research they want to do (this was never an issue for Alex, who felt he had ``so much time" to meet with groups and found his research group before classes began).  

Classes are undoubtedly the most familiar and structured aspect of the first year of graduate school, and are the only place where students receive a grade for their performance.  Thus, for students feeling unsure of how to navigate graduate school, the existing structure sends the message that focusing on classes should be their top priority, perhaps at the expense of other responsibilities like finding a group. It is therefore important for departments to reform existing graduate curricula to assure that coursework is not in tension with students' ability to find a group.

To this end, programs should consider ways of explicitly integrating students' advisor search into the graduate curriculum.  Combining these features of the first year experience would help alleviate students' perception that they have to sacrifice time looking for a group in order to focus on classes.  Moreover, revising course content to highlight available research in the department and allow students to explore potential areas of interest would more closely align with the goal of a curriculum that prepares students for their future research. 

\textbf{Clarifying funding expectations and increasing transparency of funding status are particularly important:}

Alex and Brianna's cases illustrate the critical importance of illuminating the role of funding in students' search for a group.  Alex's prior experiences gave him the insight to know that advisors need funding in order for him to be part of their group, whereas Brianna expected to be able to teach in order to receive funding.  This allowed Alex to ask about prospective advisors' funding status and motivated his application for a graduate fellowship.  Meanwhile, advisors' expectations about funding sources did not match Brianna's, which made her search far more difficult as all four astrophysics advisors told her they would not be able to support her.

More broadly, departments should consider a means of making advisors' funding status more transparent to students.  Like the first four advisors Brianna approached, her E\&M professor indicated that he did not have funding immediately. The difference was he said he would ``figure something out" for her, even if it meant teaching a bit longer, and it was unclear why the other professors could not do something similar to accommodate her.  This aspect of Brianna's story points to a general ambiguity regarding advisors' funding status that should be addressed; students must take advisors at their word that they ``don't have funding," which gives advisors power to rebuff students with little explanation.

We thank the graduate students who participated in this study.  This work is supported by NSF Award HRD-1834516.

\bibliography{main.bib} 

\end{document}